\documentclass[pdflatex,sn-nature]{sn-jnl}

\usepackage[utf8]{inputenc}
\usepackage[T1]{fontenc}
\usepackage{graphicx}
\usepackage{amsmath,amssymb,amsfonts}
\usepackage{bm}
\usepackage{tikz}
\usetikzlibrary{arrows.meta,decorations.pathreplacing}
\usepackage{booktabs}
\usepackage{textcomp}
\usepackage{xcolor}

\setcitestyle{super,open={},close={}}

\newcommand{\sqcite}[1]{ref.\spacefactor1000{}\nobreak{\setcitestyle{numbers}\cite{#1}}}

\begin{document}

\title[Nuclear mass staggering explains missing stable technetium and promethium]
{Nuclear mass staggering explains missing stable technetium and promethium}

\author*[1,2,3]{\fnm{Daiki} \sur{Nishimura}}\email{dnishimu@tcu.ac.jp}
\author[1]{\fnm{Takumi} \sur{Hasegawa}}
\author[1,2]{\fnm{Rinku} \sur{Prajapat}}

\affil[1]{\orgdiv{Department of Natural Sciences},
\orgname{Tokyo City University},
\orgaddress{\street{1-28-1 Tamazutsumi},
\city{Setagaya}, \postcode{158-8557}, \state{Tokyo}, \country{Japan}}}

\affil[2]{\orgname{RIKEN Nishina Center for Accelerator-Based Science},
\orgaddress{\street{2-1 Hirosawa},
\city{Wako}, \postcode{351-0198}, \state{Saitama}, \country{Japan}}}

\affil[3]{\orgname{Center for Nuclear Study, The University of Tokyo},
\orgaddress{\street{2-1 Hirosawa},
\city{Wako}, \postcode{351-0198}, \state{Saitama}, \country{Japan}}}

\abstract{
Technetium (Tc) and promethium (Pm) are the only elements that lack stable
isotopes in the range up to bismuth
\cite{Kowarski1950-qe,Suess1951-ky,Kohman1952-ur}, a long-standing puzzle
in physics and chemistry.
We treat this absence as a problem of selecting the lowest-mass integer
proton number $Z$ in beta-stable isobaric chains with odd mass number $A$.
Three-point mass parabolas identify two-unit jumps in the local minimum,
whereas a five-point decomposition separates the smooth quadratic component from
odd-$Z$/odd-$N$ mass staggering, defined here as the mass shift of
odd-$Z$ isobars relative to neighboring odd-$N$ isobars. With this decomposition, a fitted
bulk-plus-shell mass model reproduces the smooth trend and
shell-driven bending near magic numbers, including conventional
shell-closure skips. This model, however, does not include
odd-$Z$/odd-$N$ staggering and cannot account for the Tc and Pm skips.
The separated odd-$Z$/odd-$N$ staggering remains positive across the Tc and
Pm regions and is large enough for the lowest-mass integer-$Z$ sequence to
skip Tc and Pm. Shell-model occupation
analysis suggests that this regional staggering reflects an
orbital-dependent tensor-force monopole effect in the proton-neutron
interaction. We identify
tensor-force-driven odd-$Z$/odd-$N$ mass staggering as the origin of the Tc
and Pm skips in the odd-$A$ sequence of lowest-mass isobars.
}

\keywords{technetium, promethium, stable isotopes, beta stability,
nuclear masses, isobaric mass parabola, tensor force,
shell-model calculations}

\maketitle

\noindent
Since Mendeleev arranged the elements into a
periodic table, the regularity of the elements has been an
organizing principle of chemistry. Moseley later
showed that this order is the atomic number $Z$,
the number of protons in the nucleus \cite{Moseley1913-high-frequency}.
Yet this apparently regular sequence contains two
gaps: among the elements up to bismuth, technetium (Tc, $Z=43$) and
promethium (Pm, $Z=61$) are the only ones with no stable isotope.
The question of why two values of $Z$ disappear from the
stable-element sequence is a nuclear mass problem embedded in the periodic
table.

The absence of Tc and Pm is not caused by unusual chemistry.
Tc occupies the normal Group 7 position between niobium (Nb) and ruthenium
(Ru), and Pm lies in the lanthanide series between neodymium
(Nd) and samarium (Sm). Their chemical positions and elemental properties
are ordinary \cite{Schwochau2000-technetium,Elkina2020-promethium}.
What is missing is a stable nucleus. Stability against beta
decay is determined by the ordering of nuclear masses among isobars, nuclei
with the same mass number $A=Z+N$. For each $A$, this ordering selects the
lowest-mass member at an integer value of $Z$. The Tc--Pm problem is to
identify why this sequence of lowest-mass isobars skips two
odd-$Z$ elements.

This mass-ordering puzzle has been recognized since the early development of
shell structure and beta-stability systematics. The shell model
itself emerged from the magic numbers identified by Mayer and, independently,
by Haxel, Jensen, and Suess
\cite{Mayer1948-dc,Mayer1949-kb,Haxel1949-ta}. Kowarski, Suess, and Kohman
connected the missing elements to magic numbers, pairing,
Mattauch's isobar rule, and local odd-$A$ mass systematics
\cite{Mattauch1934-isotope,Kowarski1950-qe,Suess1951-ky,Kohman1952-ur}.
The technetium case has also been used as an educational example in terms of
stable-nuclide distributions and isobaric mass parabolas
\cite{Johnstone2017-hu}. What remains to be separated
quantitatively is how much of the Tc--Pm absence comes from the smooth
beta-stability trajectory, from shell-driven bending, and from local
odd-$Z$/odd-$N$ mass differences.

Here, we show that the lack of stable Tc and Pm isotopes is not produced by a
smooth shift of the beta-stability line alone. By combining local three-
and five-point mass relations with global mass-model comparisons, we
separate the smooth quadratic motion, shell-driven bending, and
odd-$Z$/odd-$N$ staggering in odd-$A$ chains. The Tc and Pm skips appear only
when the regional odd-$Z$/odd-$N$ staggering is included in the integer-$Z$
mass ordering, suggesting a local orbital-dependent
monopole component of the tensor force in the proton-neutron interaction rather than a universal odd-$Z$ mass
offset.

\subsection*{Integer selection in isobaric chains}

For a fixed mass number $A$, beta decay and electron capture connect isobars
with different proton numbers $Z$.  The beta-stable member of such a chain
is the lowest-mass isobar at an integer value of $Z$.  Along the odd-$A$ or
even-$A$ sequence, neighboring entries differ by $\Delta A=2$; a step with
$(\Delta Z,\Delta N)=(2,0)$ advances $Z$ by two units and skips one element.
Counting from the oxygen region gives four such odd-$A$ skips,
Ar, Tc, Ce, and Pm; the complete step count is given in Extended Data Table 1.  The
even-$Z$ skips Ar and Ce are recovered in the even-even sequence, whereas
the skipped odd-$Z$ elements Tc and Pm are not recovered elsewhere among
stable nuclides.  The remaining question is why the odd-$A$ beta-stable
sequence makes the two-unit jumps in the Tc and Pm mass regions.

\begin{figure}[htbp]
\centering
\includegraphics[width=0.98\linewidth]{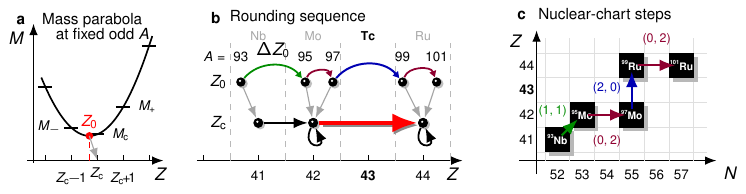}
\caption{
\textbf{| Local mass parabolas and integer-$Z$ steps.}  \textbf{a}, At a fixed odd $A$, the three neighboring masses $M_-$,
$M_{\rm c}$, and $M_+$ at $Z_{\rm c}-1$, $Z_{\rm c}$, and
$Z_{\rm c}+1$ determine a local parabola and its continuous minimum
$Z_0$.  \textbf{b}, The Tc-region example with $A=93,\ldots,101$ shows how
$Z_0$ changes as $A$ increases by two units, while $Z_{\rm c}$ is obtained
by rounding $Z_0$ to the nearest integer.  \textbf{c}, The same Tc-region steps
appear on the nuclear chart as
$(\Delta Z,\Delta N)=(1,1),(0,2),(2,0)$.  A $(2,0)$ step advances
$Z_{\rm c}$ by two units and skips one integer value of $Z$.
}
\label{fig:integer_selection_schematic}
\end{figure}

Figure~\ref{fig:integer_selection_schematic} shows the local
integer selection problem.  Panel \textbf{a} introduces the
mass-parabola minimum at fixed odd $A$, while panels \textbf{b} and \textbf{c} show how
rounding a moving continuous minimum gives
$(\Delta Z,\Delta N)=(1,1),(0,2)$, or $(2,0)$ steps in a Tc-region sequence.

The Bethe--Weizs\"acker mass formula provides the standard basis for the
isobaric mass parabola for neighboring nuclei at fixed $A$
\cite{Weizsaecker1935-kernmassen,Bethe1936-stationary}; details of the bulk
expression are given in the next subsection.  Here, this parabolic picture serves as the
standard local description of beta stability: nuclei on the neutron-rich
side of the minimum lower their mass by beta decay, whereas nuclei on the
proton-rich side lower their mass by electron capture or positron emission
($\beta^+$ decay).
The beta-stable member of the chain is the lowest-mass nucleus at
the integer value of $Z$ nearest to the minimum of the local mass parabola.

The smooth local parabola is written as
\begin{equation}
M_{\rm par}(A,Z)
=
M_0(A)+C(A)\left[Z-Z_0(A)\right]^2 ,
\end{equation}
where $M_0(A)$ represents the mass at the bottom of the local parabola, $C(A)$ denotes
the curvature coefficient, and $Z_0(A)$ indicates the continuous $Z$ coordinate of
the minimum.  The physical proton number is an integer, but
$Z_0(A)$ is generally fractional.  The continuous minimum gives the
smooth beta-stability coordinate, whereas the actual beta-stable
nucleus is selected from integer values of $Z$.  For neighboring odd-$A$
entries separated by $\Delta A=2$, we define the change in the continuous
minimum coordinate as
\begin{equation}
\Delta Z_0(A)
=
Z_0(A+2)-Z_0(A).
\end{equation}
In the ideal smooth-parabola limit, the corresponding integer proton number is
$Z_{\rm c}(A)={\rm round}[Z_0(A)]$.  Its step for $\Delta A=2$ is
$\Delta Z_{\rm c}(A)=Z_{\rm c}(A+2)-Z_{\rm c}(A)$, which corresponds on
the nuclear chart to
$(\Delta Z,\Delta N)=(\Delta Z_{\rm c},\,2-\Delta Z_{\rm c})$.
As $A$ changes by two units, the smooth minimum $Z_0$ moves by
$\Delta Z_0$, but the integer proton number $Z_{\rm c}$ changes only when
$Z_0$ crosses a nearest-integer boundary.  If two successive values of
$Z_0$ round to the same integer, then $\Delta Z_{\rm c}=0$, corresponding to
a $(0,2)$ step.  If $Z_{\rm c}$ advances by one, the step is $(1,1)$.
If $Z_{\rm c}$ advances by two, the step is $(2,0)$ and one integer value
of $Z$ is skipped.  A skipped integer value requires the continuous
minimum $Z_0$ to cross more than one nearest-integer interval during the
$\Delta A=2$ step, corresponding to $\Delta Z_0>1$ apart from exact boundary
cases.

This local construction is obtained directly from measured masses by using
decompositions around the observed lowest-mass member $Z_{\rm c}$.  The
three-point decomposition uses the masses at $Z_{\rm c}-1$, $Z_{\rm c}$,
and $Z_{\rm c}+1$ to determine a local parabola.  The extracted quantity
$Z_0^{(3)}(A)$ estimates the local minimum coordinate, while
$M_0^{(3)}(A)$ and $C^{(3)}(A)$ give the corresponding minimum-mass and
curvature quantities.  Its finite difference,
$\Delta Z_0^{(3)}(A)=Z_0^{(3)}(A+2)-Z_0^{(3)}(A)$, tracks how the
locally observed minimum moves between neighboring entries.

The five-point decomposition uses the same center $Z_{\rm c}$ but extends
the window to $Z-Z_{\rm c}=-2,-1,0,1,2$.  It writes
\begin{equation}
M^{(5)}(A,Z)
=
M_0^{(5)}(A)
+C^{(5)}(A)\left[Z-Z_0^{(5)}(A)\right]^2
+p(Z)\{S(A)+\Delta s(A)(Z-Z_{\rm c})\},
\label{eq:five_point_decomposition}
\end{equation}
where $p(Z)=+1$ for odd $Z$ and $p(Z)=-1$ for even $Z$.
This decomposition separates the smooth quadratic component from the
odd-$Z$/odd-$N$ displacement within the same local window.  The coefficient
$S(A)$ represents the symmetric odd-$Z$/odd-$N$ staggering amplitude, while
$\Delta s(A)$ represents the left-right staggering asymmetry.  The
three-point decomposition locates the observed local motion of the mass
minimum, whereas the five-point decomposition separates that motion into a
smooth quadratic part and an odd-$Z$/odd-$N$ staggering part.  The explicit
formulae for the extracted quantities are given in Methods.

\subsection*{Shell bending of beta stability}

The beta-stability trend contains a smooth component and local
nuclear-structure contributions.  The smooth component reflects the
macroscopic balance between Coulomb energy, symmetry energy, and the
proton--neutron mass difference.  A Bethe--Weizs\"acker-type bulk expression
provides a transparent reference for this trend
\cite{Weizsaecker1935-kernmassen,Bethe1936-stationary}; the fitted
bulk-plus-shell reference model is defined in Methods.  On the nuclear chart,
subtracting the fitted bulk surface leaves extended residual structures near
major shell closures, which are reduced by the fitted shell contribution
(Extended Data Fig. 1).

\begin{figure}[htbp]
\centering
\includegraphics[width=0.98\linewidth]{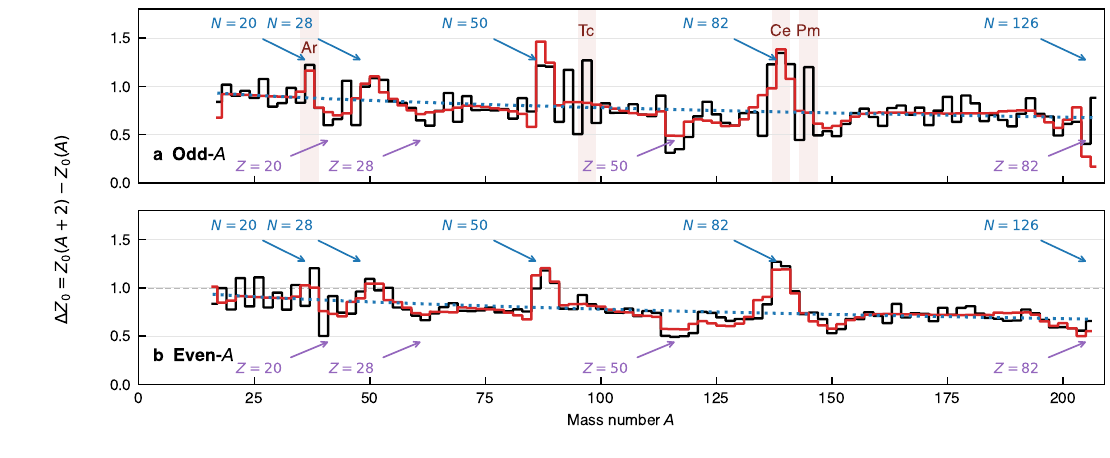}
\caption{
\textbf{| Shell bending of the beta-stability trajectory.}  The finite difference
$\Delta Z_0(A)=Z_0(A+2)-Z_0(A)$ is shown for \textbf{a}, odd-$A$ chains,
and \textbf{b}, even-$A$ chains.  The
black step curve shows the experimental three-point quantity extracted
from the adopted experimental masses, while the dotted and red curves show the same
three-point extraction applied to the bulk and bulk-plus-shell reference
mass surfaces, respectively.
}
\label{fig:principle_shell_dz0}
\end{figure}

The effect of local shell structure can be seen most directly in the
finite difference of the minimum coordinate,
$\Delta Z_0(A)=Z_0(A+2)-Z_0(A)$.
Figure~\ref{fig:principle_shell_dz0} compares the experimental
three-point quantity $Z_0^{(3)}$ with the same three-point extraction
applied to two reference mass surfaces: the smooth bulk expression alone
and the same expression after adding the phenomenological shell
contribution described in Methods.  This contribution
contains localized structures associated with the major neutron
and proton magic numbers.  It is not a microscopic shell calculation; its
purpose is to test whether broad shell-related effects can account for the
observed motion of the beta-stability trend.

The sign of this bending is naturally understood from the finite-difference
meaning of $\Delta Z_0$.  This quantity is the local change of a
minimum-coordinate curve $Z_0(A)$ between neighboring chains.  If a neutron shell
stabilizes nuclei near a particular $N=A-Z$, then increasing $A$ while
remaining near that neutron number requires an increase in $Z$.  A neutron
shell feature tends to increase $\Delta Z_0$.  Conversely,
if a proton shell stabilizes nuclei near a particular $Z$, the path is
shifted less in the $Z$ direction, and $\Delta Z_0$ tends to
decrease.

The smooth bulk reference gives a slowly varying baseline.  In particular,
it remains below $\Delta Z_0=1$ and does not by itself
produce the two-unit jump in $Z_{\rm c}$ associated with a missing element.
The shell contribution accounts for much of the broad non-smooth structure,
including the Ar and Ce regions, but it does not reproduce the Tc and Pm
anomalies in the odd-$A$ sequence.  Shell-driven bending is a necessary
part of the local motion of the beta-stability trend, but it is not
sufficient; the remaining integer-$Z$ ordering problem is tested below using
the five-point decomposition.

The results are organized around the discrete integer-$Z$ problem.
We applied five-point decomposition to an adopted experimental mass
table based on AME2020 \cite{Wang2021-fo,Huang2021-bt}, updated with recent
post-AME2020 measurements and supplemented by NUBASE2020 nuclear properties
where needed \cite{Kondev2021-is}.
The resulting data set covers 97 odd-$A$ chains; the few incomplete
five-point windows were completed as described in Methods.
First, we compare the quantities extracted from the five-point decomposition
of the experimental masses, published global mass models, and the fitted
bulk-plus-shell model.  In the following tables and figures, ``fit'' denotes this fitted
bulk-plus-shell model.  We
then ask how the separated smooth and
odd-$Z$/odd-$N$ staggering terms affect which integer value of $Z$ has
the lowest mass.  Finally, we apply this comparison to candidate skipped
steps, with special attention to Tc and Pm.

\subsection*{Five-point mass decomposition}

We applied the five-point decomposition defined above to the adopted
experimental masses, to the fitted bulk-plus-shell model used below, and to
five published global mass models: FRDM2012 \cite{Moller2016-qb}, KTUY05
\cite{Koura2005-ktuy}, HFB-24 \cite{Goriely2013-hfb24}, Duflo--Zuker DZ10
\cite{Duflo1995-dz}, and WS4 \cite{Wang2014-ws4}.  Using
the same odd-$A$ five-point windows for all the mass tables, we extracted
$M_0^{(5)}(A)$, $Z_0^{(5)}(A)$, $C^{(5)}(A)$, $S(A)$, and $\Delta s(A)$
from each table.

\begin{figure}[htbp]
\centering
\includegraphics[width=\textwidth]{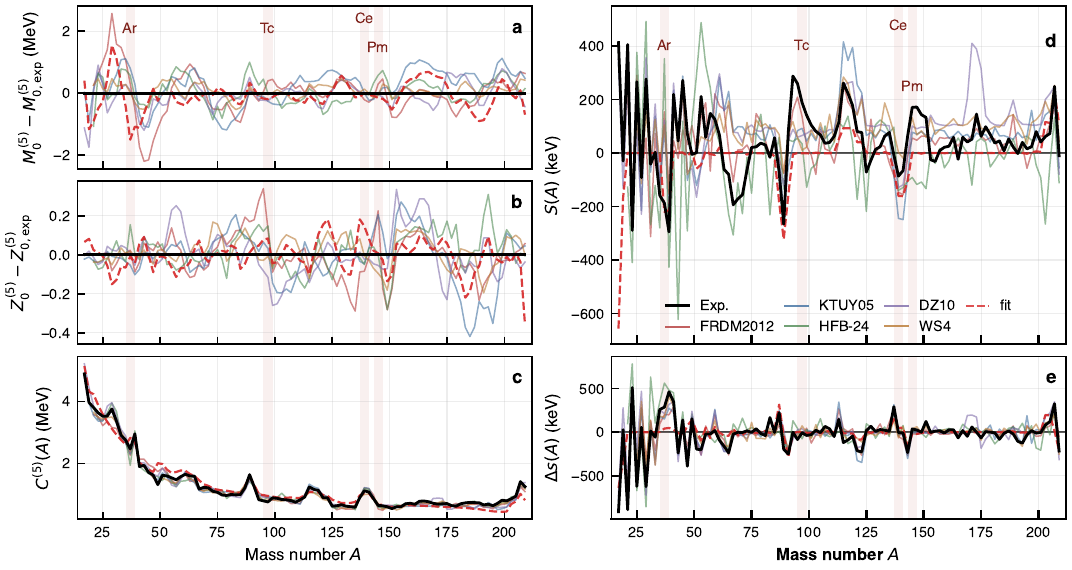}
\caption{\textbf{| Five-point decomposition of mass systematics.}  Quantities extracted from the five-point decomposition of experimental masses, five
published global mass models, and the fitted bulk-plus-shell model.  Panels
\textbf{a}--\textbf{c} show the smooth quadratic component:
$M_0^{(5)}-M_{0,\rm exp}^{(5)}$ and
$Z_0^{(5)}-Z_{0,\rm exp}^{(5)}$, and the
curvature $C^{(5)}(A)$.  Panels \textbf{d} and \textbf{e} show the separated odd-$Z$/odd-$N$ staggering
amplitude $S(A)$ and left-right staggering asymmetry $\Delta s(A)$.  The
contrast between the two columns shows why the skipped-element problem must
be tested by comparing integer-$Z$ masses rather than by comparing smooth
mass trends alone.}
\label{fig:five_point_components}
\end{figure}

Figure~\ref{fig:five_point_components} shows the $A$ dependence of
these quantities.  The left column shows the three quantities associated
with the smooth quadratic component.  The residuals in $M_0^{(5)}(A)$ and
$Z_0^{(5)}(A)$ remain small on the scale of the full mass parabola, and
$C^{(5)}(A)$ follows a common decreasing trend with shell-related structure
superposed.  With respect to these smooth quantities, the Tc and Pm regions
do not stand out as large deviations.  This distinction is important because the
skipped-element problem is determined by the ordering of neighboring integer
values of $Z$, not by these smooth quantities alone.

The right column shows the odd-$Z$/odd-$N$ staggering quantities.  In contrast to the
smooth component, $S(A)$ and $\Delta s(A)$ vary on the scale of a few
hundred keV and exhibit stronger model dependence.  The fitted shell-related
terms reproduce the sign and approximate magnitude of $S(A)$ in some regions,
but not in others.  The sign of $S(A)$ is not
universal: the experimental $S(A)$ sequence has a negative region around
$A=67$--71, whereas the Tc and Pm regions are positive.  Around shell closures,
$\Delta s(A)$ often changes from positive to negative as $A$ increases, as
seen near $A=87$--91, 137--143, and 207--209.  This sign-changing structure
follows from the way $\Delta s(A)$ compares the right and left sides of the
five-point window: a shell-stabilized nucleus enters the window
asymmetrically and changes the outer and inner right-left mass differences.  Far from shell
closures, $\Delta s(A)$ is typically smaller than the symmetric amplitude
$S(A)$.  Therefore, the relevant scale for neighboring integer-$Z$ ordering is primarily
set by $S(A)$.  Because $S(A)$ shifts odd-$Z$ and even-$Z$ masses in
opposite directions, its effect on a neighboring odd-even mass difference
is approximately $2S(A)$.  Values of $S(A)$ of order 100--200 keV can
change integer-$Z$ mass orderings when the competing masses are separated
by only a few hundred keV.

\subsection*{Skipped candidates and Tc--Pm anomaly}

We now focus on the Tc and Pm regions.  The bulk-plus-shell reference
introduced above provides the shell-modified local background,
but the remaining question is whether the observed odd-$Z$/odd-$N$ staggering
is large enough to change which integer value of $Z$ has the lowest mass.
The comparison below separates this observed staggering from the smooth
displacement of the stability trajectory.  The extracted $S(A)$ term is
used here as an empirical staggering amplitude, not as an independent mass
prediction.
For the sign convention used here, positive $S(A)$ raises the odd-$Z$
members relative to the neighboring odd-$N$ members in mass.

The preceding comparison shows that large excursions of the three-point
minimum coordinate are localized in mass number rather than global shifts of
the stability trajectory.
Table~\ref{tab:delta_z0_candidates} summarizes the three-point displacement
$\Delta Z_0^{(3)}$ for odd-$A$ candidate steps, comparing experiment,
published mass models, the fitted bulk-plus-shell model, and the same
fit after adding the empirical $S(A)$ term to the integer-$Z$ masses.  The
listed candidates are those for which the experimental
$\Delta Z_0^{(3)}$ exceeds one unit in $Z$ and at least one side has a mass
difference to a neighboring isobar below 600 keV, and for which at least one
entry in the comparison skips the candidate.  The experimental
$S(A)$ columns give the local odd-$Z$/odd-$N$ staggering scale of each step, and bold
underlined entries mark cases where integer-$Z$ mass minimization skips the
listed candidate.

\begin{table}[htbp]
\centering
\fontfamily{phv}\selectfont\scriptsize
\setlength{\tabcolsep}{3pt}
\caption{\textbf{| Skipped candidates in odd-$A$ chains.}  Three-point displacement
$\Delta Z_0^{(3)}$ for odd-$A$ candidate steps for which at least one entry in
the comparison skips the candidate.
 The columns $S_i$ and $S_f$ give the experimental odd-$Z$/odd-$N$
 staggering amplitudes for the initial and final chains, respectively.
 Bold underlined entries mark cases where integer-$Z$ mass minimization skips
 the listed candidate.}
\label{tab:delta_z0_candidates}
\begin{tabular}{lcrrrrrrrrrr}
\hline
\multicolumn{1}{c}{initial $\to$ final} & \multicolumn{1}{c}{skipped} & \multicolumn{2}{c}{$S$ (keV)} & \multicolumn{8}{c}{$\Delta Z_0^{(3)}$} \\
 & \multicolumn{1}{c}{candidate} & $S_i$ & $S_f$ & Exp. & FRDM & KTUY & HFB & DZ & WS4 & fit & fit+$S$ \\
\cmidrule(lr){1-2}\cmidrule(lr){3-4}\cmidrule(lr){5-12}
$^{139}$La$\to$$^{141}$Pr & \textbf{\underline{Ce}} & -86 & -66 & \textbf{\underline{1.35}} & \textbf{\underline{1.46}} & \textbf{\underline{1.53}} & \textbf{\underline{1.46}} & \textbf{\underline{1.39}} & \textbf{\underline{1.24}} & \textbf{\underline{1.38}} & \textbf{\underline{1.45}} \\
$^{97}$Mo$\to$$^{99}$Ru & \textbf{\underline{Tc}} & 189 & 141 & \textbf{\underline{1.27}} & 0.93 & 0.93 & 0.53 & 0.98 & 0.95 & 0.83 & \textbf{\underline{1.13}} \\
$^{37}$Cl$\to$$^{39}$K & \textbf{\underline{Ar}} & -187 & -295 & \textbf{\underline{1.22}} & 0.96 & 1.08 & 0.75 & 1.02 & \textbf{\underline{1.06}} & 1.16 & \textbf{\underline{1.28}} \\
$^{87}$Sr$\to$$^{89}$Y & Sr & -70 & -266 & 1.21 & 1.06 & 1.09 & 1.09 & 1.16 & 1.04 & \textbf{\underline{1.46}} & \textbf{\underline{1.59}} \\
$^{145}$Nd$\to$$^{147}$Sm & \textbf{\underline{Pm}} & 171 & 172 & \textbf{\underline{1.20}} & 0.64 & 0.94 & -0.21 & 1.02 & 1.03 & 0.73 & \textbf{\underline{1.16}} \\
$^{51}$V$\to$$^{53}$Cr & Cr & 1 & 212 & 1.09 & 1.02 & 1.16 & 0.57 & \textbf{\underline{1.26}} & 1.00 & 1.10 & 1.02 \\
$^{53}$Cr$\to$$^{55}$Mn & Mn & 212 & 59 & 1.07 & \textbf{\underline{1.34}} & 1.03 & \textbf{\underline{1.44}} & 1.18 & 1.05 & 0.94 & 1.04 \\
\hline
\end{tabular}
\end{table}

The table separates two effects.  The Ce skip is reproduced by all five
published mass models, consistent with a conventional two-unit step driven
by the $N=82$ shell closure.  By contrast, the Tc and Pm skips are absent
from these models, even though their experimental $\Delta Z_0^{(3)}$ values
are among the largest in the table.  The fit+$S$ column is not an independent
prediction; it tests whether the observed odd-$Z$/odd-$N$ displacement has
the sign and scale needed to change the lowest-mass integer value of $Z$.
For a neighboring odd-$Z$/odd-$N$ pair, adding $S(A)$ changes their relative
mass by about $2S(A)$, which is enough to alter the ordering when the
competing integer-$Z$ masses differ by only a few hundred keV.

The distinction between a local odd-$A$ skip and an element-level absence is
also essential.  Sr, Ar, and Ce can be skipped in an odd-$A$ sequence but
recovered by the even-even sequence, whereas Tc and Pm are odd-$Z$
entries with no such recovery.  The Mn row is a cautionary case: FRDM2012
and HFB-24 would produce a local odd-$A$ skip of Mn, although this is not the
experimental sequence.  In a full odd-$A$ scan, the fit+$S$ test reproduces
the Tc and Pm skips without introducing any additional skipped odd-$Z$
entries outside the empirical candidate set.

The Tc and Pm rows are therefore the key tests.  In the experimental
odd-$A$ sequence, Tc is skipped in the $^{97}$Mo$\to$$^{99}$Ru step and Pm
in the $^{145}$Nd$\to$$^{147}$Sm step.  In both cases, the neighboring
integer-$Z$ masses are separated by only a few hundred keV, the published
models do not produce the two-unit jump in $Z_{\rm c}$, and the fit+$S$ test
does.  This behavior is not based on a single isolated point: the
experimental $S(A)$ values remain positive over the neighboring odd-$A$
chains $A=95$, 97, 99, and 101 in the Tc region and $A=143$, 145, 147, and
149 in the Pm region.  Tc and Pm are not simply the largest smooth
displacements of the stability trajectory; they are cases in which a
persistent odd-$Z$/odd-$N$ staggering trend occurs near a degeneracy between
neighboring integer-$Z$ masses.

\subsection*{Tensor-monopole interpretation of the staggering sign}

The sustained positive sign of $S(A)$ in the Tc and Pm regions suggests
that the effect is tied to local shell and orbital structure, rather than
to a universal odd-$Z$ mass offset.  The neighboring odd-$A$ chains around
Tc and Pm already show the same positive staggering trend, so the effect is
not confined to the two chains that define the skipped steps.

Mass differences of this type are known to carry information on valence
proton-neutron interactions \cite{Brenner2006-bd,Wu2016-ab} and
orbital-dependent shell evolution
\cite{Otsuka2005-pf,Sorlin2008-zf,Otsuka2020-iq}.  The sign
of $S(A)$ provides a direct handle on such orbital-dependent effects.  To
test whether the sign of the observed staggering is compatible with a tensor-force monopole
mechanism, we constructed an occupation-based indicator
$P_{\rm tensor}(A)$ from shell-model occupation numbers; the calculation
details are given in Methods.

The construction follows the tensor-force monopole rule discussed by Otsuka
et al.: an odd nucleon in a $j_>$ orbital interacting with occupied partner
$j'_<$ orbitals, or the reverse combination, gives an attractive
contribution, whereas $j_>\otimes j'_>$ and $j_<\otimes j'_<$ combinations
give repulsive contributions \cite{Otsuka2005-pf,Otsuka2020-iq}.  Since the
present observable is a mass, attractive contributions lower the mass and
repulsive contributions raise it.  A more positive value of $P_{\rm tensor}$ thus
corresponds to a more repulsive, or less attractive, tensor-monopole
environment for the odd-$Z$ members relative to the neighboring odd-$N$
members, the same direction as positive $S(A)$.

\begin{figure}[htbp]
\centering
\includegraphics[width=\textwidth]{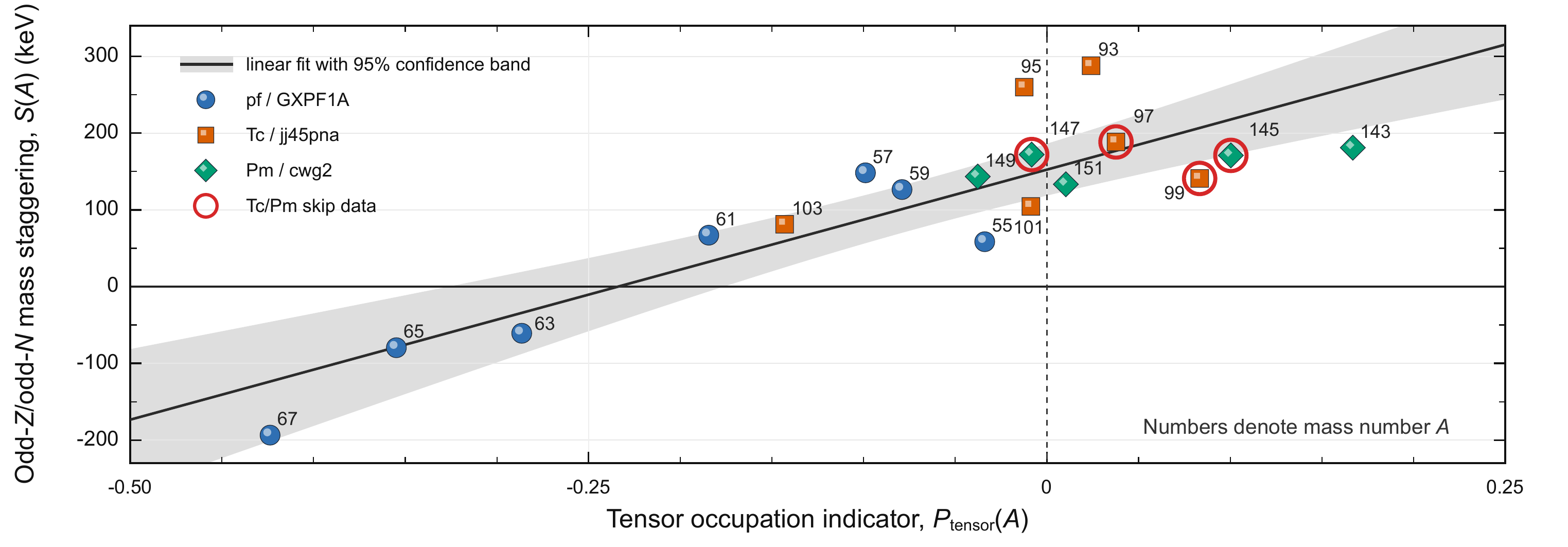}
\caption{\textbf{| Tensor-monopole indicator of odd-$Z$/odd-$N$ staggering.}  The experimental five-point staggering amplitude $S(A)$ is
plotted against the occupation-based tensor-monopole indicator
$P_{\rm tensor}(A)$.  Details of the shell-model calculations and the
construction of $P_{\rm tensor}(A)$ are given in Methods.  The line is a
least-squares fit to the 18 displayed windows, and the grey band is the
95\% confidence band.  Open red circles mark the four odd-$A$ chains
associated with the Tc and Pm skips.  Their location on the high-$S(A)$
side, relative to the negative-$S(A)$ pf-shell comparison nuclei, shows
that the observed Tc/Pm staggering has the sign expected from the monopole
effect of the tensor force in the proton-neutron interaction.}
\label{fig:tensor_monopole_proxy}
\end{figure}

Figure~\ref{fig:tensor_monopole_proxy} compares
$P_{\rm tensor}(A)$ with the experimental five-point staggering
amplitude $S(A)$.  The indicator is not intended to be a full mass
calculation; it does not include the absolute monopole energy, deformation,
or all radial matrix-element strengths.  Nevertheless, it captures the sign
separation between the Tc/Pm regions and the comparison nuclei in the
pf shell.  The Tc and Pm points lie on the high-$S(A)$ side of this
comparison, where the tensor-monopole indicator is larger than in
the negative-$S(A)$ pf-shell region.  The same construction gives different
signs in the comparison nuclei in the pf shell, indicating that the Tc and Pm
staggering is region specific rather than a universal odd-$Z$ mass offset.

We have reformulated the absence of stable Tc and Pm isotopes as a problem
of integer-$Z$ mass ordering in beta-stable isobaric chains.  The key
quantity is not only the smooth beta-stability trajectory, but the lowest
integer proton number $Z_{\rm c}(A)$ and whether it advances by two units
between neighboring odd-$A$ chains.

The three-point analysis locates where the local mass minimum moves, while
the five-point decomposition separates this motion into a smooth quadratic
component and the odd-$Z$/odd-$N$ staggering terms $S(A)$ and
$\Delta s(A)$.  The smooth component and fitted shell contribution account
for broad shell-driven bending, but not for the Tc and Pm skips.  These
skips appear when the few-hundred-keV odd-$Z$/odd-$N$ staggering is included
in the integer-$Z$ mass-ordering test.

The positive staggering in the Tc and Pm regions is not a universal odd-$Z$
offset; other mass regions show negative or sign-changing behavior.  A
tensor-monopole indicator based on shell-model occupations gives positive values for
the Tc and Pm windows, suggesting a local
orbital-dependent monopole effect of the tensor force across the $N=50$ and $N=82$
regions.  Tc and Pm are not isolated chemical anomalies,
but the element-level consequence
of discrete integer-$Z$ selection on a locally structured nuclear mass
surface.

\makeatletter
\def\@biblabel#1{#1.}
\makeatother

\section*{Methods}

\subsection*{Three-point mass-parabola quantities}

With respect to a fixed mass number $A$, three-point extraction assumes that the
three neighboring masses around the lowest-mass isobar are described locally
by a quadratic function of $Z$,
\begin{equation}
M^{(3)}(A,Z)
=
M_0^{(3)}(A)+C^{(3)}(A)\left[Z-Z_0^{(3)}(A)\right]^2 ,
\label{eq:smooth_mass_parabola}
\end{equation}
where $M_0^{(3)}(A)$ represents the minimum mass,
$Z_0^{(3)}(A)$ denotes the continuous $Z$ coordinate at the mass minimum,
and $C^{(3)}(A)$ indicates the curvature. The three-point quantities are
obtained from neighboring masses around $Z_{\rm c}$. For an odd-$A$ chain,
we use the three adjacent integer-$Z$ masses. For a given $Z_{\rm c}$, let
$M_-=M(A,Z_{\rm c}-1)$, $M_{\rm c}=M(A,Z_{\rm c})$, and
$M_+=M(A,Z_{\rm c}+1)$.
Solving for these three quantities yields
\begin{align}
M_0^{(3)}(A)
&=
M_{\rm c}-
\frac{(M_+ - M_-)^2}{8(M_- -2M_{\rm c}+M_+)},
\label{eq:three_point_m0}
\\
Z_0^{(3)}(A)
&=
Z_{\rm c}+
\frac{M_- - M_+}{2(M_- -2M_{\rm c}+M_+)},
\label{eq:three_point_z0}
\\
C^{(3)}(A)
&=
\frac{M_- -2M_{\rm c}+M_+}{2}.
\label{eq:three_point_c}
\end{align}
Along an odd-$A$ or even-$A$ sequence, the three-point step between
adjacent entries is
$\Delta Z_0^{(3)}(A)=Z_0^{(3)}(A+2)-Z_0^{(3)}(A)$.
To facilitate comparison with the odd-$A$ sequence, the same three-point decomposition
was also applied to the even--even sequence in even-$A$ chains, using masses
at $Z_{\rm c}-2$, $Z_{\rm c}$, and $Z_{\rm c}+2$. Compared with the
odd-$A$ expressions in equations~\ref{eq:three_point_m0}--\ref{eq:three_point_c},
the shift $Z_0^{(3)}-Z_{\rm c}$ is doubled and
the extracted curvature $C^{(3)}$ is divided by four, whereas the expression for
$M_0^{(3)}$ is unchanged.

\subsection*{Five-point decomposition}

Our central odd-$A$ analysis is based on a five-point decomposition of
isobaric chains.  Repeating equation~\ref{eq:five_point_decomposition} for each
odd mass number, the five neighboring masses around $Z_{\rm c}$ were represented
as
\begin{equation}
M^{(5)}(A,Z)
=
M_0^{(5)}(A)
+C^{(5)}(A)\left[Z-Z_0^{(5)}(A)\right]^2
+p(Z)\{S(A)+\Delta s(A)(Z-Z_{\rm c})\},
\end{equation}
where $Z-Z_{\rm c}=-2,-1,0,1,2$, and $p(Z)=+1$ for odd $Z$ and
$p(Z)=-1$ for even $Z$.
The five quantities $M_0^{(5)}(A)$, $Z_0^{(5)}(A)$, $C^{(5)}(A)$, $S(A)$,
and $\Delta s(A)$ were obtained by solving the five linear equations and
expressing the smooth quadratic component in completed-square form.
Let $M_{\rm c}=M(A,Z_{\rm c})$,
$M_{\pm1}=M(A,Z_{\rm c}\pm1)$, $M_{\pm2}=M(A,Z_{\rm c}\pm2)$, and
$p_{\rm c}=p(Z_{\rm c})$. The smooth quantities are
\begin{align}
M_0^{(5)}(A)
&=
\frac{-M_{-2}+4M_{-1}+10M_{\rm c}+4M_{+1}-M_{+2}}{16}
\nonumber\\
&\quad
-
\frac{(M_{-2}+2M_{-1}-2M_{+1}-M_{+2})^2}
{32(M_{+2}+M_{-2}-2M_{\rm c})},
\label{eq:five_point_m0}
\\
Z_0^{(5)}(A)
&=
Z_{\rm c}
+
\frac{M_{-2}+2M_{-1}-2M_{+1}-M_{+2}}
{2(M_{+2}+M_{-2}-2M_{\rm c})},
\label{eq:five_point_z0}
\\
C^{(5)}(A)
&=
\frac{M_{+2}+M_{-2}-2M_{\rm c}}{8}.
\label{eq:five_point_c}
\end{align}
The symmetric odd-$Z$/odd-$N$ staggering amplitude is
\begin{equation}
S(A)=p_{\rm c}\,
\frac{M_{+2}+M_{-2}-4(M_{+1}+M_{-1})+6M_{\rm c}}{16}.
\end{equation}
The left--right staggering asymmetry across the five-point window is
\begin{equation}
\Delta s(A)
=
p_{\rm c}\,
\frac{(M_{+2}-M_{-2})-2(M_{+1}-M_{-1})}{8}.
\end{equation}
The quantity $S(A)$ represents the symmetric odd-$Z$/odd-$N$ mass displacement
within the five-point window. For $\Delta s(A)=0$, positive $S(A)$ indicates
that odd-$Z$/even-$N$ nuclei are above the smooth quadratic component in mass
and even-$Z$/odd-$N$ nuclei are below it. The quantity $\Delta s(A)$ represents
the left--right change in this displacement across the five-point window,
canceling the smooth quadratic component by construction.

\subsection*{Experimental mass table}

The experimental mass table used in this work was constructed from AME2020
\cite{Wang2021-fo,Huang2021-bt}, with explicit overrides for
$^{27}$P, $^{77}$As, $^{95}$Tc, $^{133}$I, $^{189}$W, and $^{205}$Au
\cite{Yandow2023-pv,Ge2024-wl,Ge2024-bl,Valverde2024-in,Mukai2025-hg,Amanbayev2023-zi},
and for the reported preprint values of $^{179}$Yb and $^{180}$Yb
\cite{Brown2026-yb}.

The nominal dataset contains 97 odd-$A$ chains with $17 \le A \le 209$
and 97 even-$A$ chains with $16 \le A \le 208$. For odd
mass numbers, the analysis used the five isobars at $Z_{\rm c}-2$,
$Z_{\rm c}-1$, $Z_{\rm c}$, $Z_{\rm c}+1$, and $Z_{\rm c}+2$, where
$Z_{\rm c}(A)$ represents the observed lowest-mass integer proton number. For even
mass numbers, the corresponding five isobars around the even--even
member, including the odd--odd neighbors. Three chains have one missing
edge mass ($^{173}$Er, $^{174}$Er, or $^{178}$Ta); for these chains, the
local quantities were extracted from the remaining four masses
by using $\Delta s(A)=0$. All the other chains used the full five masses.
The local windows contain 967 measured mass entries among 970
nominal entries for the experimental decompositions. The same experimental
input is used for the fitted reference model below. Four-proton-unbound
nuclei, $^{16}$F, $^{16}$Ne, $^{19}$Na, and $^{39}$Sc, were retained because
measured mass values are available.

\subsection*{Global mass models}

Five-point decomposition was applied to five published global mass
models: FRDM2012~\sqcite{Moller2016-qb}, KTUY05~\sqcite{Koura2005-ktuy}, HFB-24~\sqcite{Goriely2013-hfb24}, Duflo--Zuker DZ10~\sqcite{Duflo1995-dz}, and WS4~\sqcite{Wang2014-ws4}.
These models represent different classes of global mass modeling:
FRDM2012 is a finite-range droplet macroscopic--microscopic model, KTUY05
is a spherical-basis mass formula with an improved even--odd term, HFB-24
is a Hartree--Fock--Bogoliubov model based on Skyrme and pairing
functionals, DZ10 is a microscopic mass formula, and WS4 is a global mass
formula with a surface-diffuseness correction.
For each model, mass values were obtained at the same odd-$A$ five-point
windows used for the experimental masses. The same five-point
decomposition was used to obtain $M_0^{(5)}(A)$, $Z_0^{(5)}(A)$,
$C^{(5)}(A)$, $S(A)$, and $\Delta s(A)$.

\subsection*{Tensor-monopole occupation indicator}

The indicator plotted on the horizontal axis of Fig.~\ref{fig:tensor_monopole_proxy} was
obtained from occupation numbers calculated with KSHELL~\cite{Shimizu2019-kshell}.
Related shell-model calculations have been used to study the beta decay of
$^{99}$Tc~\cite{Ramalho2024-tc99}; here we use the calculated occupations
rather than excitation spectra or transition strengths.
The construction first defines a signed occupation contribution $\xi(A,Z)$
for each calculated integer-$Z$ member, then forms the window averages
$\Xi_{\rm oddZ}(A)$ and $\Xi_{\rm oddN}(A)$ used in
$P_{\rm tensor}(A)$.
The pf-shell comparison nuclei were calculated with the GXPF1A interaction
\cite{Honma2005-gxpf1a}, the Tc-region nuclei with jj45pna
\cite{Lisetskiy2004-jj45}, and the Pm-region nuclei with cwg2
\cite{Brown2005-cwg}.  In each odd-$A$
chain, odd-$Z$/even-$N$ members were assigned an unpaired proton and
even-$Z$/odd-$N$ members were assigned an unpaired neutron.
For constructing the occupation indicator, orbitals treated as part of the
inert core by a given interaction were still included as fully occupied
partner orbitals when the relevant spin--orbit doublet was not closed.  This
convention affects only the occupation-based indicator and not the KSHELL
diagonalization itself.

For each calculated integer-$Z$ member, the orbital occupations were used to
identify the particle or hole carried by the odd nucleon.  We denote
its signed valence occupation expectation in orbital $\alpha$ by
$n_{\alpha}^{\rm val}(A,Z)$, where $\alpha$ labels the active
orbital.  The components of $n_{\alpha}^{\rm val}$ are not constrained to be integers;
they describe how the odd particle or hole is distributed over the active
orbitals.  The sign is defined relative to the nearest paired reference
configuration on the odd side, so particle and hole configurations enter with
opposite signs.  The
occupation of the partner nucleon species in orbital $\beta$ is denoted by
$n_{\beta}^{\rm partner}(A,Z)$.  Both occupations are divided by the
corresponding orbital degeneracies,
$2j_\alpha+1$ and
$2j_\beta+1$.  The
tensor-monopole sign factor for the orbital pair,
$p_{\alpha\beta}(A,Z)$, is set to $-1$ for attractive
$j_>\otimes j_<$ or $j_<\otimes j_>$ pairs and $+1$ for repulsive
$j_>\otimes j_>$ or $j_<\otimes j_<$ pairs.  The signed occupation
contribution for the member $(A,Z)$ is denoted by $\xi(A,Z)$ and is given by
\begin{equation}
\xi(A,Z)
=
\sum_{\alpha,\beta}
p_{\alpha\beta}(A,Z)
\,
\frac{n_{\alpha}^{\rm val}(A,Z)}{2j_\alpha+1}
\,
\frac{n_{\beta}^{\rm partner}(A,Z)}{2j_\beta+1}.
\end{equation}
For a given odd-$A$ chain, the same five integer-$Z$ members used in the
mass decomposition were used to construct $P_{\rm tensor}(A)$.  The odd-$Z$
and odd-$N$ members in this five-point window were averaged separately,
\begin{equation}
\Xi_{\rm oddZ}(A)=
\frac{1}{n_{\rm oddZ}}
\sum_{\substack{Z\ {\rm odd}\\{\rm window}}}\xi(A,Z),\qquad
\Xi_{\rm oddN}(A)=
\frac{1}{n_{\rm oddN}}
\sum_{\substack{A-Z\ {\rm odd}\\{\rm window}}}\xi(A,Z),
\end{equation}
where $n_{\rm oddZ}$ and $n_{\rm oddN}$ are the corresponding numbers of
terms in the five-point window.  The plotted tensor-monopole occupation
indicator is one half of the difference between the two averages,
\begin{equation}
P_{\rm tensor}(A)
=
\frac{\Xi_{\rm oddZ}(A)-\Xi_{\rm oddN}(A)}{2}.
\end{equation}

This dimensionless quantity should not be interpreted as a tensor-force matrix element or tensor energy.  It does not
include absolute radial matrix elements, deformation, or the full monopole
Hamiltonian.  It is used only as an occupation-based indicator of the sign
and relative scale expected from this monopole component.

\subsection*{Fitted bulk-plus-shell model}

We constructed a phenomenological bulk-plus-shell model as a fitted
reference model. The model is not intended to compete with global mass
models. The purpose of the model is to test which characteristics of the Tc--Pm problem are
captured by a smooth bulk trajectory plus phenomenological shell terms and
which are associated with odd-$Z$/odd-$N$ staggering separated
by five-point decomposition. The bulk binding energy and conventional
pairing term are those defined in equation~\ref{eq:bulk_binding}. The
local odd-$Z$/odd-$N$ component not generated by this conventional pairing
term is extracted separately by five-point decomposition.

The model is evaluated for given $A$ and $Z$, with $N=A-Z$.
The calculated quantity is the atomic mass excess, not the nuclear binding
energy itself. We write the free-particle contribution in
mass-excess units as follows:
$M_{\rm free}(A,Z)=Z M_{\rm H}+(A-Z)M_{\rm n}$, where $M_{\rm H}$ represents the
hydrogen-atom mass excess and $M_{\rm n}$ represents the neutron mass excess.
This convention includes the electron mass through $M_{\rm H}$; additional
electron binding energies of heavier neutral atoms are not included explicitly.
The bulk binding energy is
\begin{equation}
\begin{split}
B_{\rm bulk}(A,Z)
=&\;
a_{\rm v}A
-a_{\rm s}A^{2/3}
-a_{\rm c}Z(Z-1)A^{-1/3}
-a_{\rm a}\frac{(A-2Z)^2}{A} \\
&\;
+a_{\rm ex}Z^{4/3}A^{-1/3}
+\delta(A,Z).
\end{split}
\label{eq:bulk_binding}
\end{equation}
Here $a_{\rm v}$, $a_{\rm s}$, $a_{\rm c}$, and $a_{\rm a}$ are the
volume, surface, Coulomb, and quadratic asymmetry coefficients. The
coefficient $a_{\rm ex}$ represents a Coulomb-exchange-like correction of
the type used in droplet and finite-range droplet mass formulae
\cite{Myers1969-average,Moller1995-frdm,Moller2016-qb}. The conventional
pairing term is $\delta(A,Z)=+a_{\rm p}A^{-1/2}$ for even-$Z$/even-$N$
nuclei, $\delta(A,Z)=0$ for odd-$A$ nuclei, and
$\delta(A,Z)=-a_{\rm p}A^{-1/2}$ for odd-$Z$/odd-$N$ nuclei.
The complete mass expression is
\begin{equation}
M_{\rm bulk+shell}(A,Z)
=
M_{\rm free}(A,Z)
-B_{\rm bulk}(A,Z)
-\Delta B_{\rm shell}(A,Z),
\label{eq:shell_midshell_mass}
\end{equation}
where $\Delta B_{\rm shell}$ represents the contribution of the phenomenological shell to
the binding energy.

The shell contribution is expressed as the sum of magic-number Gaussian terms
and phenomenological midshell profiles in terms of the neutron and proton number,
\begin{equation}
\begin{aligned}
\Delta B_{\rm shell}(A,Z)
={}&
\sum_i h_i
\exp\left[-\frac{(N-N_i)^2}{2\sigma_i^2}\right]
+
\sum_j h_j
\exp\left[-\frac{(Z-Z_j)^2}{2\sigma_j^2}\right]
\\
&-W^{(n)}(N)-W^{(p)}(Z).
\end{aligned}
\label{eq:shell_midshell_correction}
\end{equation}
The midshell boundaries are
$N_i=8,20,28,50,82,126$ for neutrons and
$Z_j=8,20,28,50,82$ for protons. Among these entries, the active
Gaussian centers in equation~\eqref{eq:shell_midshell_correction} are
$N_i=50,82,126$ and $Z_j=28,50,82$.

For a given neutron number $N$, we choose the interval that contains it.
If $N_i \le N \le N_{i+1}$, the boundary magic numbers of this interval are
$N_i$ and $N_{i+1}$, and the corresponding midshell depths are
$d_i^{(n)}$ and $d_{i+1}^{(n)}$.  The neutron midshell profile in this
interval is then
\begin{equation}
W^{(n)}(N)
=
\{(1-t_n)d_i^{(n)}+t_n d_{i+1}^{(n)}\}
\frac{t_n^{q_i}(1-t_n)^{q_{i+1}}}
{\mathcal{N}_i^{(n)}},
\label{eq:smoothed_midshell}
\end{equation}
where $t_n=(N-N_i)/(N_{i+1}-N_i)$. This definition gives $t_n=0$ at the
lower magic number and $t_n=1$ at the upper magic number. The factor
$\mathcal{N}_i^{(n)}$ normalizes only the dimensionless shape factor
$t_n^{q_i}(1-t_n)^{q_{i+1}}$ in this interval; the interpolated depth remains
set by $d_i^{(n)}$ and $d_{i+1}^{(n)}$.
The proton profile is defined by the same expression with $N$, $N_i$,
$N_{i+1}$, $d_i^{(n)}$, $d_{i+1}^{(n)}$, and $\mathcal{N}_i^{(n)}$
replaced by $Z$, $Z_j$, $Z_{j+1}$, $d_j^{(p)}$, $d_{j+1}^{(p)}$, and
$\mathcal{N}_j^{(p)}$. The profile is joined continuously at
magic-number boundaries. When the neutron or proton coordinate is outside
the range near $^{16}$O or $^{209}$Bi, it is reflected at the nearest
magic number before $W^{(n)}$ or $W^{(p)}$ is used.
The sign convention in
equation~\eqref{eq:shell_midshell_mass} indicates that positive
Gaussian amplitudes increase the binding and lower the mass near magic
numbers, whereas positive midshell profiles lower the binding and raise the
mass between magic numbers.
Because these terms are fitted phenomenologically, $\Delta B_{\rm shell}$ may
also absorb local mass variations associated with deformation and other
structure effects; it should be considered as an effective shell-structure
term rather than a pure magic-number correction.

The model parameters were determined by a least-squares fit to the
five-point quantities extracted from the experimental mass table.
The fit used 194 local chains, which contain 967 measured masses in the
corresponding five-point windows. The minimized objective was
\begin{equation}
\begin{aligned}
\chi^2
=
\sum_A
\left[
\left\{\frac{\Delta M_0^{(5)}(A)}{0.5~{\rm MeV}}\right\}^2
+
\left\{\frac{\Delta Z_0^{(5)}(A)}{0.1}\right\}^2
+
\left\{\frac{\Delta C^{(5)}(A)}{0.2~{\rm MeV}}\right\}^2
\right],
\end{aligned}
\label{eq:shell_midshell_chi2}
\end{equation}
where the $\Delta$ symbols denote model-minus-experiment residuals of the
corresponding five-point quantities.
To ensure comparable contributions from the three residual classes, the scale factors
(0.5 MeV, 0.1, and 0.2 MeV) were iteratively selected to approximate the standard
deviations of the corresponding residuals.
The fitted bulk-plus-shell model has 34 free parameters: five independently
fitted bulk coefficients and 29 midshell and Gaussian coefficients. The
Coulomb-exchange-like coefficient was limited by the droplet model
ratio $a_{\rm ex}/a_{\rm c}=0.7636$ used in FRDM-type mass formulae, so that
the exchange term remains a standard small correction to the leading
Coulomb-asymmetry surface rather than an additional free shape parameter. No
residual for the odd-$Z$/odd-$N$ staggering amplitude $S(A)$ or the
left--right staggering asymmetry $\Delta s(A)$ was included in the objective
function. Both quantities obtained from the fitted model are modeled
outputs, not directly fitted quantities.

The numerical
parameters and the resulting root-mean-square residuals for the fitted
five-point quantities are listed in Extended Data Table 2. To directly
validate the mass surface, the root-mean-square residual of the calculated
atomic mass excess over the 967 measured masses is 2.220 MeV when the shell
term is switched off and 0.925 MeV when it is included. For a broader validation,
the corresponding residual over all 2114 measured masses with
$17\leq A\leq 209$ in the table is 2.074 MeV. The largest residuals
in this broader validation occur for light neutron-rich nuclides; no mass
with $A\geq 60$ has $|\Delta M|>5$ MeV. These validations confirm that this
parameter set is tuned to the present local five-point analysis rather than
to a global reproduction of the nuclear chart. As a
complementary integer-$Z$ validation, choosing the lowest mass in each five-point
window identifies the observed $Z_{\rm c}$ in 173 of the 194 local chains
(89.2\%): 78 of 97 odd-$A$ chains and 95 of 97 even-$A$ chains.

The fitted model is subsequently analyzed with the same five-point decomposition used in
the experimental masses and global mass models. The extracted quantities
$M_0^{(5)}(A)$, $Z_0^{(5)}(A)$, $C^{(5)}(A)$, $S(A)$, and $\Delta s(A)$
have the same definitions for all mass tables.

\section*{Data availability}

The mass data used in this study are available from public nuclear mass
tables. The processed tables generated in this study will be made available
upon reasonable request.

\section*{Code availability}

The analysis scripts used in this study will be made available upon
reasonable request.

\section*{Author contributions}

D.N. conceived the study, developed the mass-decomposition and
tensor-occupation analyses, wrote the analysis codes, performed the
calculations, prepared the figures and wrote the manuscript. T.H. contributed
to the formulation of the physical problem, independently checked the
analysis and interpretation, and revised the manuscript. R.P. investigated
relevant literature, contributed to the interpretation of odd-$Z$/odd-$N$
staggering and tensor-monopole effects, checked the manuscript logic and
language, and revised the manuscript. All authors discussed the results and
approved the final manuscript.

\section*{Competing interests}

The authors declare that they have no competing interests.

\section*{Use of generative AI}

During the preparation of this work, the authors used OpenAI's ChatGPT and
Codex to assist with manuscript drafting, language refinement, code
inspection, and consistency checks. The authors reviewed and edited all
AI-assisted output and take full responsibility for the scientific content,
calculations, figures, references, and conclusions.

\section*{Additional information}
Correspondence and requests for materials should be addressed to D.N.
Reprints and permission information is available at www.nature.com/reprints.

\section*{Extended Data}

\noindent\textbf{Extended Data Table 1 | Counting of $\Delta A=2$ steps.}
Counting of $\Delta A=2$ steps in the beta-stable sequences considered in
the main text.  The even-$A$ row represents the even-even branch counted from
$^{16}\mathrm{O}$ to $^{208}\mathrm{Pb}$, excluding light-mass exceptions.
The odd-$A$ row is counted from $^{17}\mathrm{O}$ to $^{209}\mathrm{Bi}$.
Because pairing favours even-even nuclei over odd-odd nuclei, the even-$A$
beta-stable branch in this mass region contains no $(1,1)$ steps.  The four
$(2,0)$ steps in the odd-$A$ sequence correspond to Ar, Tc, Ce, and Pm.
Ar and Ce have stable isotopes in the even-$A$ sequence, leaving Tc and Pm as
the elements in this range that lack stable isotopes.

\begin{center}
\fontfamily{phv}\selectfont\small
\begin{tabular}{c c c c c c c c}
\toprule
Sequence
& Range
& Intervals
& $\Delta Z_{\rm total}$
& $\Delta N_{\rm total}$
& $(2,0)$
& $(1,1)$
& $(0,2)$ \\
\midrule
Even-$A$
& $^{16}\mathrm{O}$--$^{208}\mathrm{Pb}$
& 96
& 74
& 118
& 37
& 0
& 59 \\
Odd-$A$
& $^{17}\mathrm{O}$--$^{209}\mathrm{Bi}$
& 96
& 75
& 117
& 4
& 67
& 25 \\
\bottomrule
\end{tabular}
\end{center}

\vspace{1.5em}

\begin{center}
\includegraphics[width=\linewidth]{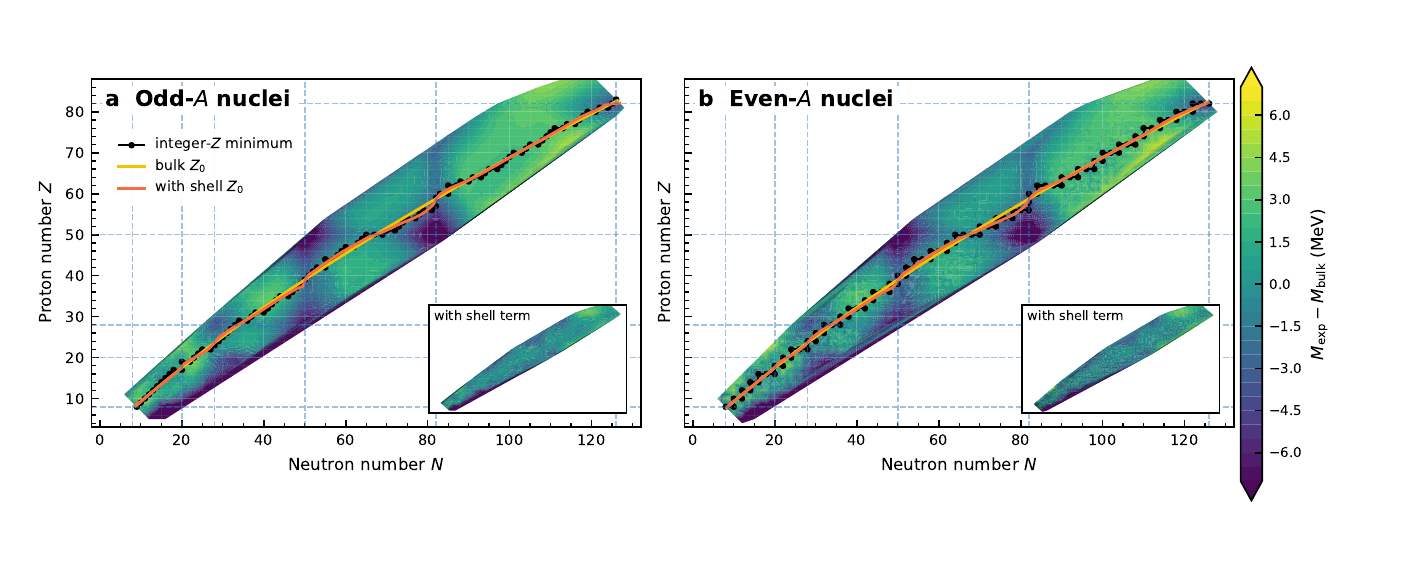}
\end{center}

\noindent\textbf{Extended Data Fig. 1 | Bulk residuals and fitted shell correction.}
Odd-$A$ nuclei and even-$A$ nuclei on nuclear charts of
$M_{\rm exp}-M_{\rm bulk}$.  The even-$A$ panel contains both even-even
and odd-odd nuclei.  Black points and lines show the integer-$Z$ minima,
while the yellow and orange curves show three-point minimum coordinates
extracted from the bulk and bulk-plus-shell reference mass surfaces,
respectively.  The inset in each panel shows the remaining residual after
the shell contribution is included.

\vspace{1.5em}

\noindent\textbf{Extended Data Table 2 | Parameters of the fitted bulk-plus-shell model.}
Adopted parameters of the fitted bulk-plus-shell mass model.  Energy
coefficients $a$, Gaussian amplitudes $h$, midshell depths $d^{(n)}$ and
$d^{(p)}$, and fit root-mean-square residuals in $M_0$ and $C$ are given in
MeV; residuals in $Z_0$ are dimensionless.  Gaussian widths $\sigma$ are in
units of neutron number or proton number.  The ratio $a_{\rm ex}/a_{\rm c}$
is fixed to 0.7636.

\begin{center}
\fontfamily{phv}\selectfont\small
\begin{minipage}[t]{0.45\linewidth}
\centering
\textbf{Bulk parameters}

\vspace{0.2em}

\begin{tabular}{lr}
\toprule
Parameter & Value \\
\midrule
$a_{\rm v}$ & 15.159662 \\
$a_{\rm s}$ & 16.98044 \\
$a_{\rm c}$ & 0.679233 \\
$a_{\rm a}$ & 21.0251 \\
$a_{\rm ex}$ (fixed) & 0.51867 \\
$a_{\rm p}$ & 16.999 \\
\bottomrule
\end{tabular}
\end{minipage}
\hfill
\begin{minipage}[t]{0.50\linewidth}
\centering
\textbf{Magic-number Gaussian parameters}

\vspace{0.2em}

\begin{tabular}{lrr}
\toprule
Center & \multicolumn{1}{c}{$h$} & \multicolumn{1}{c}{$\sigma$} \\
\midrule
$N_i=50$  & 4.088 & 6.461 \\
$N_i=82$  & 4.393 & 3.517 \\
$N_i=126$ & 3.353 & 3.540 \\
$Z_j=28$  & 2.709 & 2.637 \\
$Z_j=50$  & 3.596 & 4.184 \\
$Z_j=82$  & 1.086 & 2.234 \\
\bottomrule
\end{tabular}
\end{minipage}

\vspace{0.8em}

\begin{minipage}[t]{0.50\linewidth}
\centering
\textbf{Midshell parameters}

\vspace{0.2em}

\begin{tabular}{rrrr}
\toprule
Boundary & $d^{(n)}$ & $d^{(p)}$ & $q$ \\
\midrule
8   & 2.627 & $<0.001$ & 0.050 \\
20  & 1.750 & 0.017 & 0.940 \\
28  & 5.042 & 0.916 & 1.719 \\
50  & 4.724 & 1.332 & 0.870 \\
82  & 1.209 & 3.570 & 0.774 \\
126 & $<0.001$ & -- & 8.107 \\
\bottomrule
\end{tabular}
\end{minipage}
\hfill
\begin{minipage}[t]{0.43\linewidth}
\centering
\textbf{rms residual}

\vspace{0.2em}

\begin{tabular}{lr}
\toprule
Quantity & Value \\
\midrule
odd $\Delta M_0$ & 0.469 \\
odd $\Delta Z_0$ & 0.083 \\
odd $\Delta C$ & 0.198 \\
even $\Delta M_0$ & 0.503 \\
even $\Delta Z_0$ & 0.085 \\
even $\Delta C$ & 0.236 \\
\bottomrule
\end{tabular}
\end{minipage}
\end{center}

\end{document}